\documentclass[pre,twocolumn,nofootinbib]{revtex4-1}
\usepackage[utf8]{inputenc}
\usepackage{graphicx}
\usepackage{xcolor}
\newcommand{\correction }[1]{ #1 }
\begin{document}
	\title{Pinning of crack fronts by hard and soft inclusions: a phase field study }
	\author{Hervé Henry}
	\affiliation{Laboratoire  PMC, Ecole Polytechnique, CNRS, IP Paris, 91120 Palaiseau Cedex}
	\begin{abstract}
		Through tridimensonal numerical simulations of crack propagating in material with an elastic moduli heterogeneity it is shown that the presence of a simple inclusion can affect dramatically the propagation of the crack. Both the presence of soft and hard inclusions can lead to the arrest of a crack front. Here  the mechanism leading to the arrest of the crack are described and shown to depend on the nature of the inclusion. This is also the case in regimes where the presence of the inclusion leads to a slow down of the crack.
	\end{abstract}
	\maketitle

\section{Introduction}
  While most theoretical work on fracture propagation has been devoted to the study of fracture propagation in homogenous materials, brittle or ductile, most materials that are actually used are heterogeneous. This is the case at  the macroscopic scale for construction materials such as stones or concrete. This is also true for materials such as nacre or polymer blends that are homogeneous at a macroscopic scale but present heterogeneities at the microscopic scale. In all cases, the heterogeneity of the microstructure leads to heterogeneities of both fracture energy and elastic moduli. When considering the elastic properties of intact materials, the later are the only one at play and homogeneization techniques have allowed to propose effective material properties, even in cases where the phase constitutive of the material have dramatically different properties as it is the case for instance  with porous materials\cite{porous2,Dormieux2002,porous1}. 
  
   However, in the context of fracture  the process zone is most of the time at a scale that is smaller than a typical representative volume element. As a result  contrarily to what happens with elastic properties,  the use of effective material properties when describing crack propagation cannot be satisfactory. This is examplified by Nacre\cite{ABID2018385,Hutchinson2001,Okumura2001}. Hence a proper description of the crack front in an heterogeneous material is first needed. In this case the possible effects of the microstructure are many. For instance, heterogeneities, by modifying the stress field locally can affect crack front trajectory. Fracture toughness heterogeneities  can lead to crack deflection or arrest and to the propagation of cracks along weak interfaces. 
   
  In this context many studies have been devoted to the effects of regions with different  fracture energy and the presence of interfaces\cite{Evans1989,He1991,He1994,He1996,HOSSAIN2014,LEBIHAIN2020,Schneider2016}. The effect of a misfitting  inclusion (i.e. with eigenstrain) on a crack has also been  studied\cite{Herrmann1995,Bhargava1977,Bower1991}. However little attention has been paid on changes  elastic moduli which can vary by order magnitude and can strongly affect crack propagation as is evidenced by the deflection of cracks toward voids\cite{Erdogan1974,Brescakovic2022} and deflection by hard inclusions\cite{Patton1990,Gao1991}. {Recently, some progress  on such materials has been made through numerical simulations using phase field models\cite{Brach2019,Clayton2014,Lacondemine2017}. The focus of these work was on the nature of the inclusion itself and on its elastic properties. For instance in \cite{Clayton2014} the inclusion is lying exactly in the plane of proapagation of the crack front and the effects of both fracture energy and material stiffness are  considered. Other studies were focused on 2D geometries while the bridging mechanism dicussed in\cite{Bower1991} and the results of \cite{LEBIHAIN2020} show clearly that 3D effects must be taken into account.  }   

  Here  the possible effects of simple heterogeneities on crack propagation are shown and  and the emphasis is put on three dimensional effects since the microstructure of most materials are  tridimensional. {More specifically this work focuses on the effects of the position of the inclusion  with respect to the crack propagation plane. In addition, since  in composite materials, the toughenning effects of inclusions are often sought in the context of catastrophic failure where the dynamic aspects of fracture propagation cannot be  overlooked\cite{Freund1991}, this work brings some light on these aspects while being limited to non branching cracks\cite{henry-08}.}

 To this purpose, a phase field model is  used and the effect of an inclusion with a different elastic moduli  on a crack front is studied. Simulations results indicate that both hard and soft inclusions affect  crack propagation and can eventually stop a crack front \correction{that is called pinning here}. However, the way the crack front is stopped by the inclusion differ. In addition in all cases, if the stress is increased the crack front can resume normal propagation without any additional crack nucleation as would be needed in two dimensional systems. The mechanisms that lead to this depinning are found to  depend strongly on the nature of the inclusion:hard with higher elastic moduli  or soft with lower elastic moduli.

\section{Model, system description  and numerics}
 \begin{figure}
\centerline{	\includegraphics[width=0.45\textwidth]{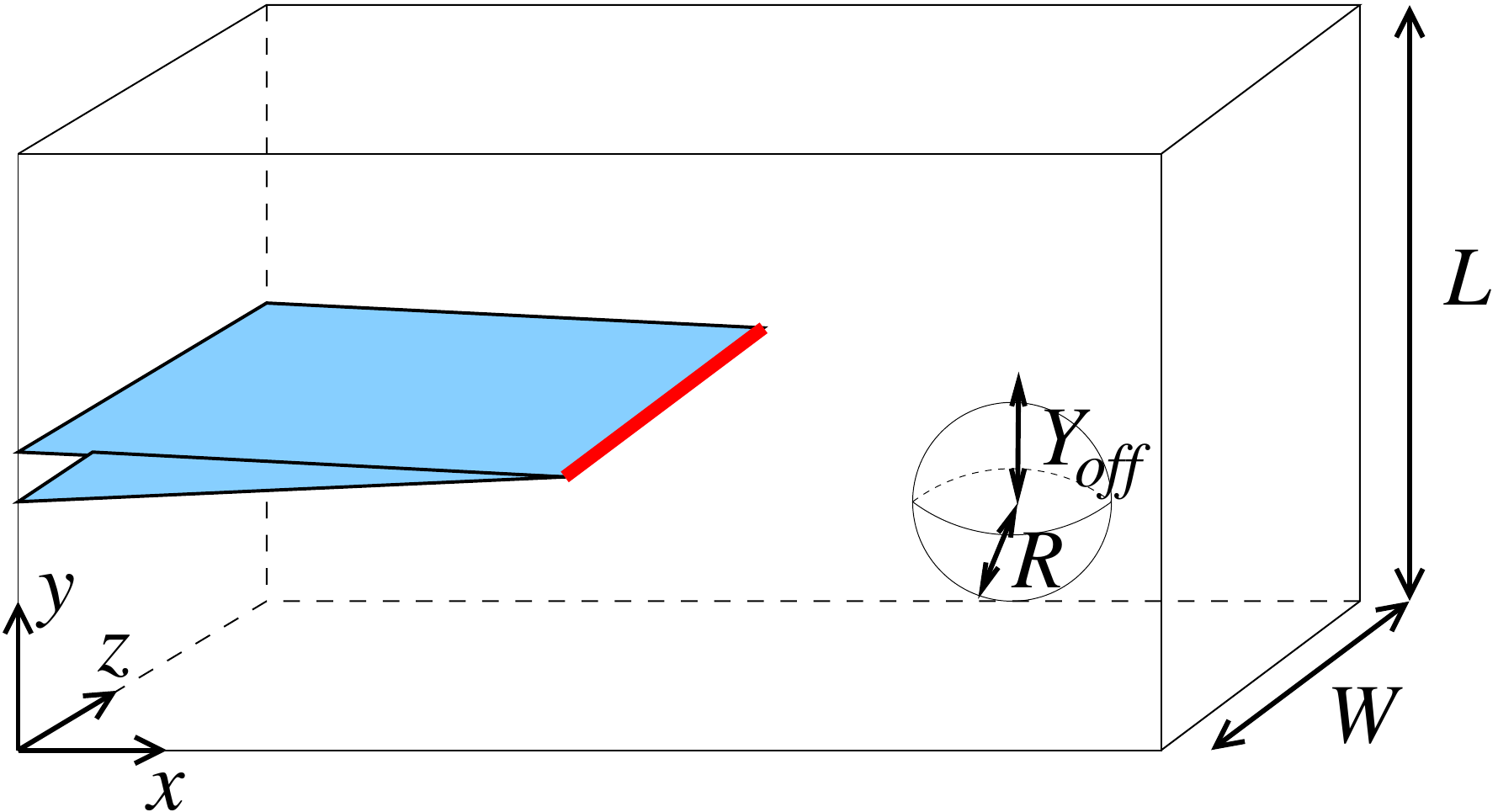}}
	\caption{\label{fig_setup} Perspective view of the simulation setup. The crack is propagating fron left to right and the inclusion is the sphere seen on the r.h.s. of the simulation domain.}
\end{figure} 
The system considered here consists of a simple spherical inclusion embedded in a matrix where a crack is propagating along the $x$ axis with an applied strain along the y axis as schematized in fig.\ref{fig_setup}.{This would correspond to a void or a softer phase in an homogeneous material such as glass or, for instance, a tougher/harder  phase in a polymeric material such as carbon black}. 
For the sake of simplicity and to avoid to have to deal with crack front curvature\cite{Bazant79, Henry_2010}, the boundary conditions along the crack front direction, denoted $z$, is periodic (in all simulations presented here, the period is 30 space units (SU), while the size of the sample in the x and y directions is 480 SU and 240 SU). The loading is imposed through the use of  fixed displacement $\Delta Y$  at top and bottom boundaries along the y axis in an infinite strip geometry.{ In the context of dynamic crack propagation (see pages 243-247 of \cite{Freund1991}), this leads to a crack propagating with a steady velocity that is function of the available elastic energy denoted $G=(0.5 \lambda+\mu) ( \Delta Y/L)^2$.   }

 Hence the system geometry is given by its length along the $y$ axis denoted $L$, its width (or period) along the z axis $W$. The radius of the spherical inclusion is denoted $R$ and the  distance between its center and the midplane of the system is denoted $b$. When $R$ is larger than $b$ the inclusion intersects the midplane that corresponds to the plane of steady crack propagation. { The crack  is initiated by the introduction of a pre crack that corresponds to a spatial distribution of the phase field with a thickness of the same order of magnitude as the regularization length. It must be noted that the crack propagates freely over a significant distance before its path is affected by the inclusion}. Simulations performed using twice smaller values of $L$, keeping $W=30$ lead to quantitatively similar results. 

To model crack propagation a phase field model is used.  It relies on the introduction of an additional variable the phase field $\varphi$ that can be seen as $1-d$ where $d$ is a damage variable. The evolution of the phase field derives from a free energy functional  that writes:
\begin{widetext}
\begin{eqnarray}
	\mathcal{F}&=&\int\left( (h V(\varphi)-\varepsilon_c g(\varphi)+\frac{D}{2}|\nabla \varphi|^2 )+ g(\varphi)\mathcal{E}_{el}\right) dV\\
  \mathcal{E}_{el}&=&\frac{\lambda(\mathbf{x})}{2}(\mathrm{tr} \epsilon)^2+\mu(\mathbf{x})\mathrm{tr}(\epsilon^2)
\end{eqnarray}
\end{widetext}
where $h=0.25$, $D=4.$ and $\varepsilon_c=0.25$ are model parameters and $\mathcal{E}_{el}$ is the elastic energy density for an intact material as a function of the strain tensor $\epsilon$. Its expression involves the space dependant coefficients $\lambda(\mathbf{x})$ and $\mu(\mathbf{x})$ that encode the material heterogeneity as will be described below. $V(\varphi)-\varepsilon_c g(\varphi)$ is a tilted  double well potential with $V=\varphi^2(1-\varphi)^2$ and $g(\varphi)=4 \varphi^3-3\varphi^4$ is a coupling function chosen so that when $\varphi=1$ the contribution of the elastic strain corresponds to the intact material while when $\varphi=0$, $g=0$ and the contribution of strain to the energy of the system is 0. It has been shown in \cite{kkl}  that with this model the equilibrium solutions in 1D (local minimas of $\mathcal{F}$) can be either $\varphi=1$ and uniform strain or localized solutions where $\varphi$ is equal to 1 except in a small region of thickness $w_\varphi$, {the so called regularization length}, where it goes toward small values ($\ll 1$).{ If, as it was the case here,  the shape of the potential  (i.e. the ratio $h/\epsilon_c$) is kept constant the regularization length scales like $D/h$. With the values considered here, $w_\phi$ is of the order of 6. SU if $w_\phi$ is the thickness of the region where $\varphi<0.5$ }.  In this case the strain is concentrated in this region and the stress in the material is vanishing\cite{kkl}. The corresponding fracture energy is $2\Gamma=1.046$ energy units and will be called in the following $G_c$. This model does not present any strain softening\cite{Kuhn2015} and has the peculiarity that contrarily to what is observed in other approaches such as the Francfort-Marigo variational approach\cite{Francfort1998,Bourdin2008,Miehe2010}, there is no crack nucleation. Hence this model is limited to the modelisation of crack propagation.   This limitation comes with the advantage that no unphysical crack nucleation can occur when running macroscopic simulations and  using a model  interface larger than the microscopic process zone.{ Indeed it has be shown that if the regularisation length used in the model is chosen larger than the actual size of the process zone, then the material stress at failure is reduced proportionally,  leading to the possibility of unphysical nucleation events at low stress values. \cite{Maurini2011,Tanne2018a}.}

In order to describe the evolution of a crack the relaxation  evolution equation\footnote{\correction{ More complex kinetic equations\cite{Jou2005} may be used, especially for fast crack propagation, however in the context of crack propagation the complexity of the events taking place in the process zone would  not allow to estimate them.}  } that derive from this free energy is :
\begin{equation}
	\partial_t\varphi=-\frac{1}{\beta} \frac{\delta\mathcal{F}}{\delta \varphi}\label{eq:relaxation}
\end{equation}
where $1/\beta$ is kinetic parameter that must be  positive or 0. It has for value $\beta_0$ here except in two cases:
\begin{itemize}
    \item if the r.h.s of eq. \ref{eq:relaxation} is positive then $1/\beta$ is 0. This ensures irreversibility.
    \item   $1/\beta$ is  $\max (0, (1/\beta_0)(A+g'(\phi)K_{\mbox{Lam\'e}}(tr \epsilon)^2)/A)$  with
      $A=-\delta \mathcal{F}/\delta \phi$ if $\mathrm{tr} \epsilon<0$ so that the
compression energy does not contribute to crack growth. However this does not prevent crack faces interpenetration.
\end{itemize}
 These two cases do not occur during normal crack propagation in the setup desribed here (mode I loading). Then, as has been shown in \cite{henry-08}, the energy dissipation  at the crack tip is proportional to $\beta_0$, taken equal to 4 here,  and the total energy of the system is guaranteed to decrease due to fracture propagation:
\begin{equation}
  \partial_t \mathcal{F}=\int \frac{\delta\mathcal{F}}{\delta \varphi}\partial_t \varphi=-\frac{1}{\beta}\int (\frac{\delta\mathcal{F}}{\delta \varphi})^2=-{\beta}\int (\partial_t \varphi)^2
\end{equation}

In addition, due to the double well potential with minima in $\varphi=0\mbox{ and }1$ and the zero slope of $g(\varphi)$ at $\varphi=0\mbox{ and }1$, the phase field cannot take values out of the 0-1 range. This model has been proven to reproduce well the Griffith criterion\cite{henry-08,kkl}.

The evolution equation for the elastic field is the wave equation\correction{ that derives from the Newton's law} with an additional damping term:
\begin{eqnarray}
  \partial_{tt}{u_i}=-\frac{\delta F}{\delta u_i}-\eta \partial_t u_i
\end{eqnarray}
where $u_i$ is the displacement field  in the $i$ direction with respect to a reference configuration.  The damping term $\eta=0.02$ is chosen large enough to prevent wave reflexions at the  boundaries of the sample. It is also chosen small enough to ensure that the wave equation is not overdamped. Here, the chosen value of $\eta$ corresponds to a damping length of the order of 100.SU for an elastic wave, to be compared with the size of the process zone of 3.SU and typical obstacle sizes of 20SU. \correction{ This term must be seen as a numerical trick to keep crack propgagation slow enough.}.

In this work the effect of spatial heterogeneity is limited to elastic properties of the material. To this purpose space dependant Lamé coefficients $\lambda(\mathbf{x})$ and $\mu(\mathbf{x})$ are considered.  They are considered to obey the following law when considering an inclusion of radius $R$~:
\begin{eqnarray}
  \lambda(\mathbf{x})&=&\lambda_1 (h(r))+\lambda_0(1-h(r))\\
   \mu(\mathbf{x})&=&\mu_1 (h(r))+\mu_0(1-h(r))\\
   h(r)&=&(\mathrm{tanh}(\frac{ r-R}{ w})+1)/2
\end{eqnarray}
where $r$ is the distance from the center of the inclusion and $\mu_{0,1}$, $\lambda_{0,1}$ are the Lame coefficients inside and outside the inclusion. Here for the sake of simplicity, $\lambda_1=\mu_1=1$ and  only two cases are presented:$\lambda_0=\mu_0=5$ for a hard inclusion and $\lambda_0=\mu_0=0.1$ for a soft inclusion.{This corresponds to variation of the Young's modulus, keeping the poisson ratio constant. Obviously other kind of heterogeneities can be considered\cite{Lacondemine2017} and may lead to different behaviour. Here, the choice has been to focus more on geometric aspects than on the nature of inclusions.} The parameter $w$ sets the abruptness of the elastic moduli change {and  the smooth variation of elastic moduli used here allows to have inclusions defined implicitly on a regular cartesian grid}. It is chosen to be 0.3 SU. Simulations with $w$ ranging between 0.3 and 1.2 SU do not lead to noticeable differences, indicating that as long as $w$ is much smaller than the characteristic size of the inclusion, no significant effects of its value   should be observed.

The model equations are solved using a finite difference scheme for spatial derivatives that ensures that the discretized equations derive from a discretized free energy. An explicit {Verlet\cite{Verlet} scheme is used for time stepping using a discretized elastic energy in order to ensure mechanical  energy conservation (without the small dissipation)} .The grid is following the crack tip so that its center is kept close to the center of the grid along the x axis (a tread-mill geometry)\footnote{{periodically the crack front (for each $z$ the point with the highest $x$ for which $\phi=0.5$) is computed and the grid is shifted by an integer number of grid points so that the whole front is behind the middle of the grid along the x axis}}. The use of a simple numerical scheme and geometry  allows to use efficiently massively parallel architecture such as GPUs. The in house code (using openACC) is used on a single GPU architecture using either an NVIDIA Tesla P100 with double precision) or a NVIDIA RTX3080 or A40  with single precision. In all instances the simulations lead to the same results indicating that the use of single precision is not detrimental here. Most of the simulations presented here have been  performed using single precision. 

\section{Results}
  \begin{figure}
  	\centerline{\includegraphics[width=0.5\textwidth]{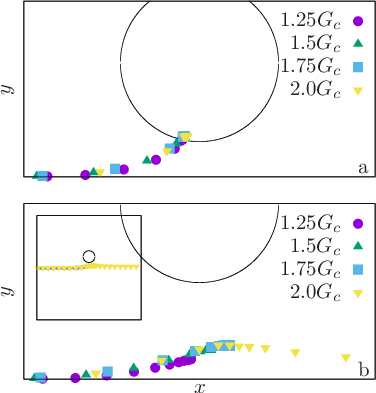}}
	  \caption{\label{fig_2d}{Plots of the crack tip  trajectories together with the inclusion boundary for 4 different values of the loading in 2D. In both cases the radius of the obstacle is 13.5, and the value of $Y_{off}$ is 19.5 in (a)   and 22.5  in b. The crack is propagating from left to right along the x axis. In the inset, the  full simulation domain is represented along the y axis and the rightmost half along the x axis. For  $Y_{off}=19.$, further increasing the load does not lead  the  crack to resume propagation. }}
  \end{figure}
  
 Here the effects of both hard and soft inclusions are described.  Since they differ dramatically both cases are discussed in two separate sections  First when the inclusion is soft , second when the inclusion is hard.    
\subsection{Soft inclusions}
  Before discussing the three dimensional case, it is useful to give a comprehensive view of the effects of a two dimensional soft inclusion on crack propagation. As is expected from the behavior of a crack in the presence of a void, the soft inclusion attracts the crack  since its softness  favors the built up of the singularity at the crack tip.  {This was for instance exemplified  in \cite{Gao1991}(p 1674-1676) where it is shown that the apparent  SIFs are higher when a softer region lies ahead of the crack}. As a results the crack deflects toward the inclusion.   Depending on the load and on the geometry of the system, different behaviors can be observed.
  
  { First, if the crack path is sufficiently far from the inclusion, then  two cases arise: if  the load is sufficient, then the crack will be  deflected toward the inclusion (its tip goes away from the middle of the sample) and slow down. Thereafter, once it has passed the vicinity of the inclusion,  it  retrieve normal propagation and its tip position relaxes toward the center of the sample. If the load is not sufficient, the crack will stop at some point on the path. It should be noted that in this case  the crack path does not intersect the inclusion, is mostly determined by the geometry and is not changed significantly by the load. This behavior are illustrated in fig.\ref{fig_2d} for $G=1.5G_c$, $G=1.75G_c$ and   $G=2,0G-c$ where the position of the crack tip are plotted at regular time intervals together with the inclusion.  For such values of $G$, the crack velocity ranges from  $0.13 \mbox{ to }0.36 \ c_s$ where $c_s$ is the shear wave speed (the Raileigh wave speed is $\approx 0.91 c_s$ here)\footnote{Using eq. 5.4.9 of \cite{Freund1991}, one can estimate the dissipation at the crack tip induced by the $\beta$ and additional dissipation term. For $G=1.25G_c$ it is $0.238 G_c$, for  $G=1.5G_c$ it is $0.446G_c$ and for $G=1.75G_c$ it is $0.669G_c$. }. Hence here, the  effect of changing $G$ is to change the way the crack goes along a \textit{prescribed path}: for low values of $G$ it slows down and eventually stops and for higher values it slows downs and eventually   retrieves normal propagation. One should note that once the crack has stopped, increasing $G$ will make advance and eventually retrieve normal propagation. }.
  
   Second, when the inclusion is closer to the crack path, independently of the load, the behavior is dramatically different. The crack deflects toward the inclusion and propagates until it crosses its boundary. There it stops{ as can be expected from \cite{Gao1991} where it is shown that the stress intensity factor at a crack tip in  in a soft inclusion is much smaller than the stress intensity factor in the bulk.} In this case, when the load is increased, once the crack tip has entered the inclusion, it will not resume propagation unless a crack nucleation occurs, for instance at  the boundary of the inclusion. 
   
\begin{figure}
 	\centerline{\includegraphics[width=.5\textwidth]{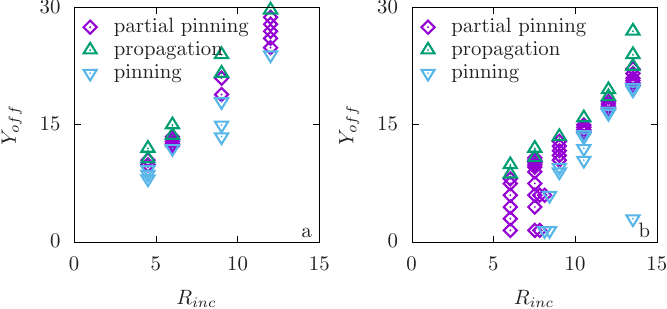}}\centerline{\includegraphics[width=.5\textwidth]{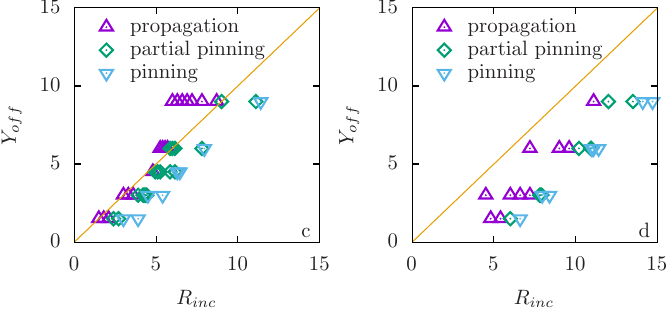}}
	  \caption{Phase diagram in the $R$, $Y$  plane for simulations for  soft inclusions  (2D in a and  3D in b) and hard inclusions (2D in c and 3D in d).  G is set to $1.5 G_c$ for soft inclusions and to $1.25 G_c$ for hard inclusions.  The criteria chosen to differentiate between freely moving is arbitrary. Here it is chosen to consider that the later was observed when the minimal crack front velocity is smaller than 1/2 its free  propagation speed.{Obviously, changing the criteria would lead to different shape of the partial pinning region.}
 		\label{fig_diagphase} }
 \end{figure}

   {
   The boundary between the different regimes, for a given value of $G=1.5G_c$ are represented on fig. \ref{fig_diagphase}. It should be noted that in this phase diagram,there is no distinction between the two kinds of crack  arrest. This choice is due to the fact that this phase diagram serves the purpose of comparing 2D and 3D results.  For the later, as will be seen, such a distinction is not relevant. }

 \begin{figure}
 	\centerline{\includegraphics[width=.45\textwidth]{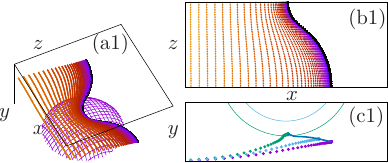}}\centerline{
 \includegraphics[width=.45\textwidth]{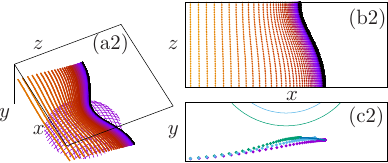}	
 }
 	\caption{\label{fig_chronograph}(a1):perspective view of the crack front and inclusion at different times  during the pinning of the crack. The crack is propagating from left to right and the color of the line is changed  according to the time.(b1) top view of the crack front. Only half of the front is shown. The side of the sample is at the bottom of the graph. (c1) side view of the position of the crack tip in different $xy$ planes (center of the sample, side of the sample and an intermediate plane). Graphs use the actual aspect ratio. a2, b2, c2~: same views in a case where the crack front is pinned but does not intersect the inclusion as it is evidenced in c2}
 \end{figure} 
 
 In the 3D case, the fact that the inclusion does not occupy the full width of the sample induces different behavior. First, as in 2D it can induce the arrest of the crack front. However, while in 2D, the crack is simply stopped, here,  the crack front is pinned on the inclusion and outside of the soft region the crack front is deformed  as has been observed experimentally when the crack front encounters a tougher region \cite{Dalmas2009}. This is exemplified in fig. \ref{fig_chronograph} (b1) where one can see a time series of crack front seen from above, a perspective view(a1)  and some cuts ($xy$ planes) of the crack path(c1). One can see that the crack front is curved and that the crack surface is bent toward the inclusion.  In this situation the crack front can be \textit{divided} into three parts: a central part where the crack front has reached the inclusion and the two side parts  where the crack front is in the homogeneous domain and cannot advance due to the pinned part of the crack front. { This behaviour is similar to the  pinning mechanism described in \cite{GaoRice1989} and was also discussed in \cite{Bower1991} where it was  was proven to contribute to  toughening for cracks confined in a plane.}  
  As in 2D there exists a situation where the crack front is stopped while it is not intersecting the inclusion at all as it is illustrated in fig. \ref{fig_chronograph} (a2, b2 and c2). In these cases, the front cannot be described as three parts. 
  	
\begin{figure}
 	\centerline{\includegraphics[width=.45\textwidth]{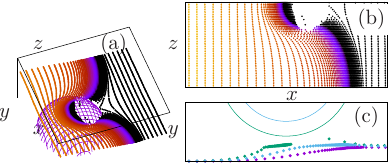}}\centerline{ \includegraphics[width=.45\textwidth]{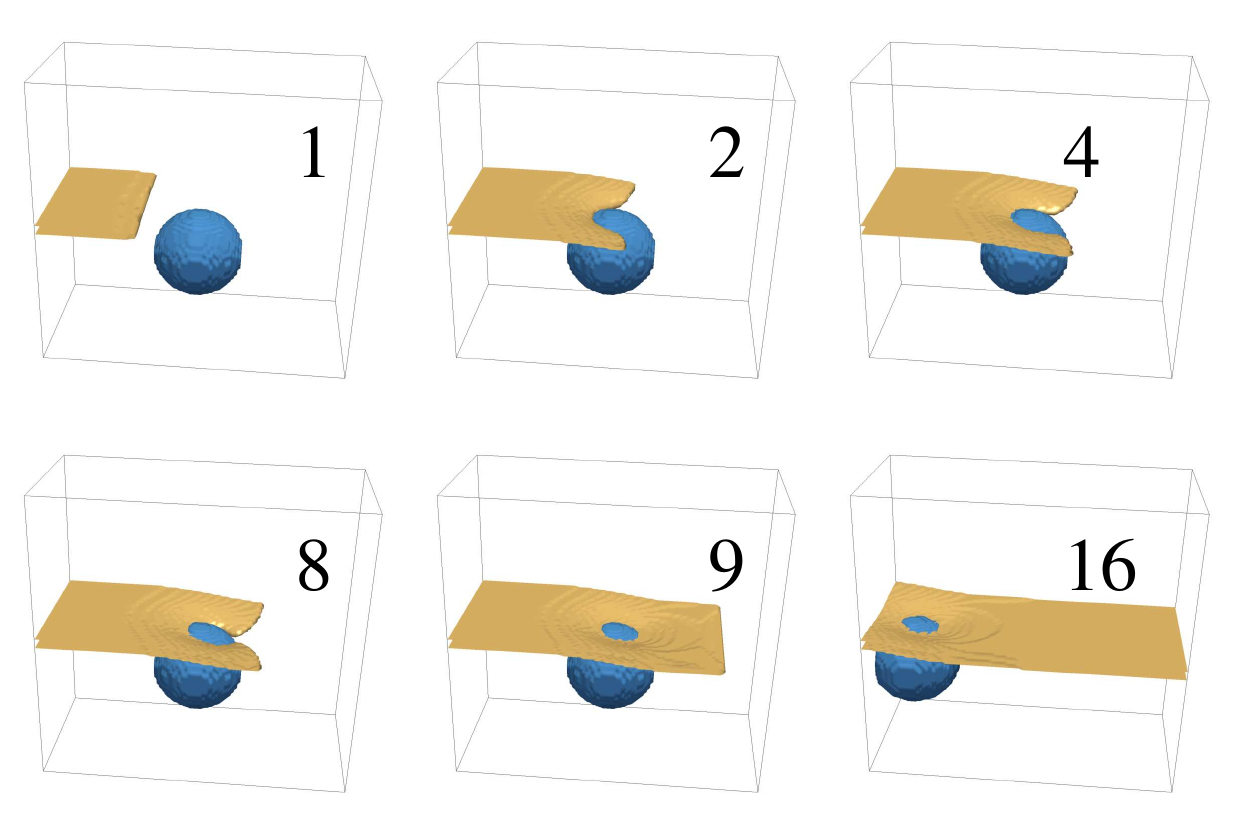}}
 	\caption{\label{fig_unpinning}\textbf{Top} (a):perspective view of the crack front and inclusion at different times  during the slow down and unpinning of the crack. The crack is propagating from left to right and the color of the line is changed  according to the time.(b) top view of the crack front. Only half of the front is shown. The side of the sample is at the bottom of the graph. (c) side view of the position of the crack tip in different $xy$ planes (center of the sample, side of the sample and an intermediate plane).\textbf{Bottom} Sequence of images of the crack (iso surface$\phi=0.5$ of the phase field) when it goes around the inclusion. The images are labeled by a dimensionless time variable. One can see that between the images 2 and 8, the crack moves very little. This contrasts with the advance of the crack between images 1 and 2 or 8 and 9. It is worth mentioning  that {the crack isosurface may intersect the sphere while the crack tip is less than a regularization length away from the obstacle} }
 \end{figure}

 In both these pinning situations (\ref{fig_chronograph} (c1) and (c2)), the normal propagation of the crack can be retrieved by simply increasing the load (once the crack has stopped) without crack nucleation. While this  is in contrast with the 2D case where, if the crack front intersect the boundary of the inclusion, a nucleation event is needed to retrieve normal crack propagation, it can be expected. Indeed, since the inclusion does not span the whole sample, there are always parts of the crack front that do not intersect the inclusion and for which the \textit{SIF reduction} induced by the inclusion is absent. For such parts they will obviously advance as the load is increased and the front motion observed in partial pinning situations (\correction{that is described below and illustrated in fig. \ref{fig_unpinning}}) is expected.

  When considering either higher loads, smaller inclusions or inclusions further apart from the midplane, an almost \textit{unperturbed} crack propagation is observed. This regime is not shown, but at intermediate values of load there exists a third regime,\correction{that is called partial pinning here}, that is worth mentioning and describing. It consists of a pinning of the crack at the inclusion followed by a slow propagation around it and finally a normal crack propagation. This is illustrated in figure\ref{fig_unpinning}(a) and (b) where the position of the crack front at different equally spaced times is plotted in perspective and from above (only half the simulation is shown there). One can see that the spacing between the front line is dramatically reduced indicating a slow down of the crack front. It is then divided into 3 regions. One in the center is pinned at the inclusion. The other two at the boundary of the samples are stopped by curvature and by the fact that since the inclusion is not broken the singularity cannot fully develop. However they  advance slowly  around the inclusion and once their \textit{free ends} meet the crack resumes normal fast propagation. Thereafter, the inclusion remains intact \correction{and forms a ligament that lies behind the crack path}  for a small time and eventually breaks when the crack front has advanced enough. This behavior is also illustrated in figure \ref{fig_unpinning} where the crack surface (the isosurface $\phi=0.5$) is plotted at different times together with the spherical inclusion. Once again, it is interesting to note that this behaviour is similar to the one discussed in \cite{Bower1991} since it implies that for at least some time, intact inclusions will remain behind the crack acting as \textit{bridges} and participating potentially to the toughening of the material.\correction{ It must be noted that such ligaments are reminiscent of patterns that can be seen  during the rupture of some elastomers\cite{ligament} at small scales. }

  Before turning to the results in the case of an hard inclusion we represent in figure \ref{fig_diagphase} the different observed behavior for a given load and different sizes of the inclusion and offset  of its center from  the plane of propagation of the crack. One can see that there is a minimal radius below which no pinning of the crack can be observed and solely a slow down (that can be important) is possible. This differs significantly from the 2D case where the transition line between   pinning and propagation can be extrapolated down to a 0 radius of the inclusion.  This, together with the results described above illustrates clearly the difference  between a 2D and a 3D inclusion.

\subsection{Hard inclusions}
  
 \begin{figure}
	\centerline{\includegraphics[width=.45\textwidth]{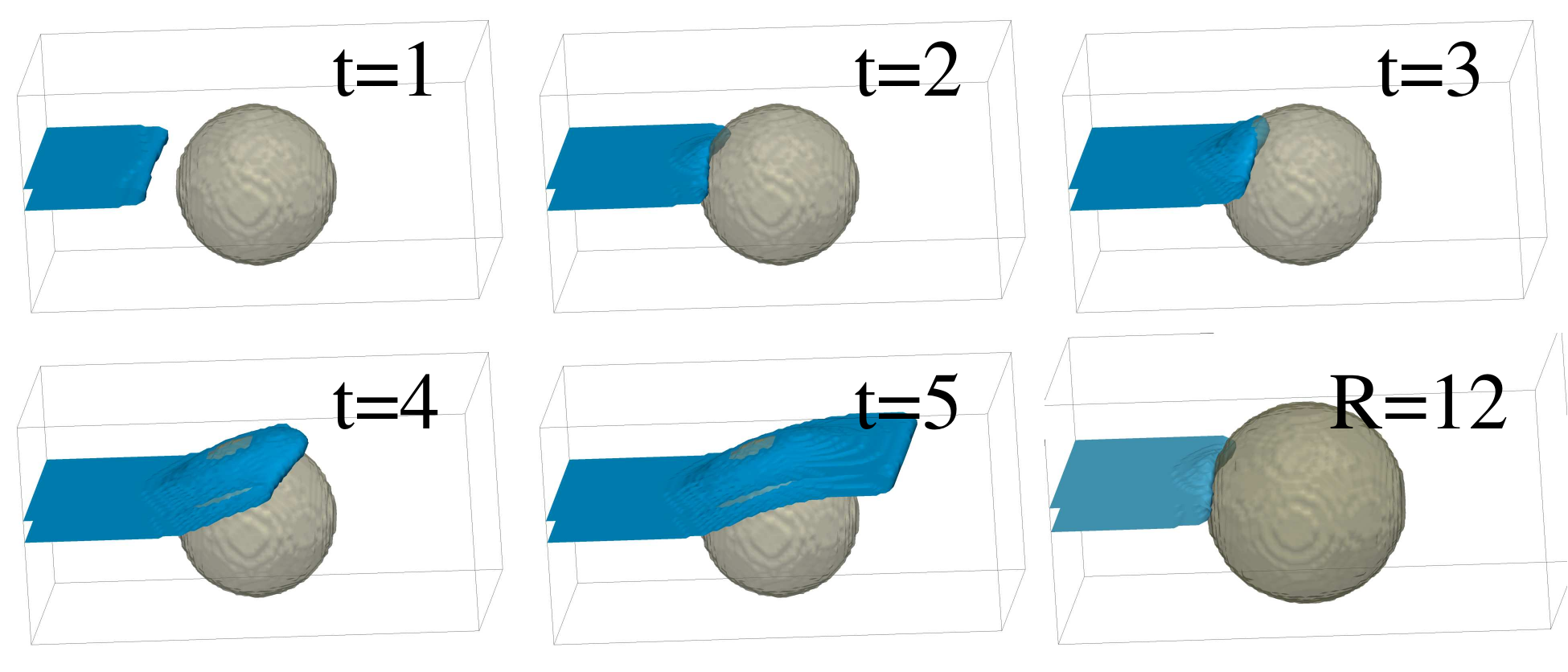}}
	\caption{\label{figsnaphard}Snapshots of the crack going around an hard  inclusion at different time of propagation The crack surface (isosurface $\phi=0.5$) is represented in blue while the spherical inclusion is in gray. At t=4 and t=5 one can see that the crack breaks partly the inclusion. The lower rightmost image corresponds to a crack stopped by an hard inclusion. In all cases, only a portion of the simulation domain is represented. }
\end{figure}

  We now turn to the case of hard inclusions and proceed similarly. First in 2D systems, 
 the inclusion tends to repel the crack. Indeed, the presence of an hard inclusion tends to prevent large strains in the surrounding material. As a result crack propagation away from the inclusion, in region with nominal elastic moduli,  is favored. As a result the crack is repelled. {This can, again, be understood using \cite{Gao1991} where it was shown that when the material ahead of the crack tip is tougher, the SIF is smaller.}
 
  However, the hard inclusion  still can pin the crack. Indeed, when  the crack tip gets too close to the inclusion, the  higher elastic moduli prevents the singularity at the crack tip to be strong it  may lead to the arrest of crack propagation. In this situation, unlike with soft inclusions,  the crack never crosses the boundary of the inclusion. As a result, simply increasing the load will lead to crack propagation without the nucleation of a new crack. 
 
  In 3D, a spherical hard inclusion can still lead to the arrest of a crack, in a similar manner: most of the crack front is stopped at the inclusion as  as is shown  fig.\ref{figsnaphard} on the bottom right snapshot (R=12).   In cases where the inclusion is  smaller or more off the crack propagation plane, the crack front will go around the inclusion with a significant slow down.   This is examplified  in fig.\ref{figsnaphard}. Once the crack front is close to the inclusion, it is stopped (T=2) and starts to propagate slowly around it (t=3, t=4). Once the inclusion is no longer ahead of the front, it retrieves normal propagation (t=5). Similar behavior can be seen with smaller inclusions that do not span the whole system. A phase diagram for a fixed value of $G=1.25G_c$ is presented in fig. \ref{fig_diagphase}. It must be noted that the inclusion will affect significantly  the crack propagation only if it lies in its path. Moreover, if the center of the spherical inclusion lies in the close vicinity of the crack path it is dramatically more efficient at pinning the crack. 
  {The contrast between the \textit{local} effect of hard inclusion and the \textit{non local } effect of soft inclusions lies in the fact that the hard (resp. soft) inclusion tends to repel (resp. attract) the crack, as a result the crack is further away from  (resp. closer to ) the inclusion and is less (resp. more) affected by the inclusion. }

  This  shows that the  difference in elastic moduli can be sufficient to prevent crack propagation through an hard inclusion and that it can lead to the arrest of the crack. {However, it must be noted that it appears to be less efficient. This may be related to the fact that in the case of hard inclusions, they deflect cracks. Indeed it has been shown that crack deflection\cite{LEBIHAIN2020}  is less efficient that pinning or bridging\cite{Bower1991,GaoRice1989}.} It must also be noted that when  there is no crack arrest, the presence of the inclusion can still  affect significantly crack propagation. Indeed, the crack front propagates around it and its propagation velocity  can dramatically decrease.

\section{Conclusion and perspective}
 Here we have  shown that, when considering crack propagation in heterogeneous materials,  elastic effects alone  can have a dramatic effect on crack propagation even when  at a macroscopic scale the heterogeneity is barely noticeable when considering the \textit{outer} stress field.  The results of the numerical simulations show that in both the cases of a soft and of an hard inclusion, it is possible to pin the crack at the inclusion. However the way the crack is affected  differ qualitatively in  two aspects. First  in the case of a soft inclusion, the effects can be significant  when the inclusion is outside of the crack path while in the case of hard inclusions, it is not the case. Second, the unpinning mechanism are dramatically different. In the case of the soft inclusion, the crack front advances, moves around it in two parts that eventually reconnect resume normal propagation  and finally once the crack front is sufficiently far away from the inclusion it eventually breaks,{ as it has  already been discussed in \cite{Bower1991} in the case of toughness heterogeneities for a confined crack front}. In the case of the hard inclusion, the crack fronts keeps it connectivity and the crack surface goes around the inclusion.  In both cases, it must be noted that the unpinning mechanism does not involve  the nucleation of a new crack contrarily to what is expected in 2D systems.
 
 This works gives some insight on the effects of heterogeneities on crack propagation. It paves the way to further work that would aim at considering more complex systems such as the effects of either disordered or crystalline arrangements of spherical inclusions, more complex inclusions (elliptical, non convex) and also of different kind of inclusions (with varying poison ratio, eigenstrain). Such studies should help creating architectured materials with higher fracture energy\cite{metamaterial}. In addition, the dramatic slow down of a crack that is observed here, may be of importance for statistical models of crack front dynamic in a quenched disorder that are used to model earthquakes\cite{Rosso2014}.  

 Another extension of this work that is also needed is to consider the effects of fracture toughness heterogeneities. However, in this case, a simple description like the one presented here is unlikely to be successful\cite{HENRY2019} and more complex approaches should be considered\cite{Nguyen2015,Nguyen2016,Schoeller2023}.  
% \bibliographystyle{unsrt}
 
% \bibliography{biblio}

\end{document}